\documentclass[aps,prl,twocolumn,groupedaddress]{revtex4}
\usepackage{epsfig}

\begin{document}

%\preprint{}

\title{Properties of compatible solutes in aqueous solution}
\author{Jens Smiatek$^1$}
\thanks{}
\email{jens.smiatek@uni-muenster.de}
\author{Rakesh Kumar Harishchandra$^2$}
\author{Oliver Rubner$^1$}
\author{Hans-Joachim Galla$^2$}
\author{Andreas Heuer$^1$}
%\homepage[]
%\thanks{1}
%\altaffiliation{Condensed Matter Theory}
\affiliation{$^1$Institut f{\"u}r Physikalische Chemie, Westf{\"a}lische Wilhelms-Universit{\"a}t M{\"u}nster, 48149 M{\"u}nster, Germany \\
$^2$ Institut f{\"u}r Biochemie, Westf{\"a}lische Wilhelms-Universit{\"a}t M{\"u}nster, 48149 M{\"u}nster, Germany}

\begin{abstract}
We have performed Molecular Dynamics simulations of ectoine, hydroxyectoine and urea in explicit solvent. Special attention has been spent
on the local surrounding structure of water molecules.
Our results indicate that ectoine and hydroxyectoine are able to accumulate more water molecules than urea by a pronounced ordering due to hydrogen bonds.
We have validated that the charging of the molecules is of main importance resulting in a well defined hydration sphere. The influence of a varying salt concentration is also investigated.
Finally we present experimental results of a DPPC monolayer phase transition that validate our numerical findings.
\end{abstract}

\date{\today}

\maketitle
\section{Introduction}
\begin{figure}[h!]
 \includegraphics[width=0.5\textwidth]{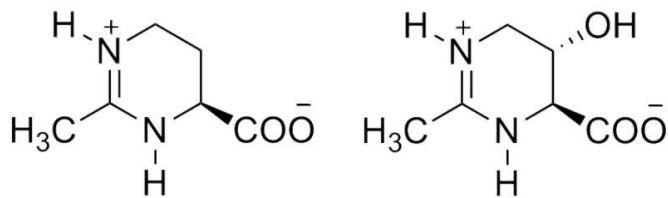}
\caption{Structural formulae of ectoine (left) and hydroxyectoine (right).
}
\label{fig:1}
\end{figure}

%\section{\label{1}}
Extremolytes are natural compounds which are synthesized by extremophilic microorganisms. Chemically they are organic osmolytes which are build of amino acids, betain, 
sugar and heteroside derivatives \cite{Lentzen06}. Thus the name extremolyte is a coinage: organic osmolytes which are
synthesized by extremophilic microorganisms. 
The presence of these molecules allows microorganisms to resist extreme living conditions like drastic temperature variations and high salinity \cite{Lentzen06,Driller08}.
Interestingly, these solutes are biologically inert and accumulate at high concentration in the cytoplasm without interfering with the overall cellular functions; hence they are 
called {\em compatible solute} \cite{Morris78}.\\ 
Long-known compatible solutes are ectoine and hydroxyectoine occurring in anaerobic chemoheterotrophic and halophilic/halotolerant bacteria \cite{Morris78,Galinski85,Galinski93,Lapidot88}.
The main characteristics are given by a low molecular weight and strong water binding properties.
It was indicated by experiments \cite{Schuh85, Lapidot93} and ab-initio molecular orbital studies \cite{Nagata98} in aqueous solution that the zwitterionic structure is more stable than the neutral form.
The structural formulae of these molecules are presented in Fig.~\ref{fig:1}.\\
Over the last years the stupendous characteristics of these small molecules have succeeded to explore a novel application area in dermatological industry. 
Long used hygroscopic molecules like urea have been systematically replaced
by extremolytes in a stepwise manner \cite{Dirschka08}.
Specifically ectoine and hydroxyectoine have been used as cell 
protectant in skin care \cite{Driller08,Dirschka08} due to their accessibility of large scale production \cite{Lentzen06,Galinski85,Galinski93,Lapidot88}. \\
Further biological importance is given by the stabilization of proteins in presence of compatible solutes \cite{Santos03,Leatherbarrow98}. 
The reason for this protective behavior can be explained by a complex interplay between all species. Several studies have investigated this mechanism in detail \cite{Lee81,Arakawa83,Arakawa85,Timasheff2002, 
Street2006, Bolen2008, Schellman2003}. 
Most of the theories prefer an indirect mechanism in which the compatible solute does not directly interact with the macromolecule. An important theory in this context is 
the preferential exclusion model \cite{Lee81,Arakawa83,Arakawa85,Timasheff2002} 
which states that the appearance of the co-solvent leads to thermodynamic interactions with the protein \cite{Timasheff2002}.
Due to these interactions, the protein repels the co-solvent from its surface to the bulk region. Thus the concentration of the co-solvent is lower in close vicinity to the protein whereas its bulk value is increased.
This can only be accomplished by the addition of excess water molecules to the proteins surface which results in a preferential hydration of the macromolecule \cite{Timasheff2002}.
The excess water molecules finally conserve the native form such that unfolding becomes less favorable which is realized by an increase of the melting temperature \cite{Lee81,Yu04,Yu07,Knapp99,Yancey81}.\\
Numerical investigations have further identified a significant decrease in the activity coefficient of water in presence of hydroxyectoine and ectoine \cite{Sadowski10}.
In addition it was found that compatible solutes fluidize lipid monolayers by acting along the line tension \cite{Galla10, Galla11}. Direct interactions have to be also taken into account as it was shown for 
high concentrations of hydroxyectoine \cite{Galla11} which again emphasizes the indirect mechanism described above. 
This effect has been shown as advantageous for cell membranes due to an acceleration of repair mechanisms and signaling processes \cite{Driller01}. 
To summarize, the presence of compatible solutes allows proteins as well as cell membranes to resist environmental 
stress and to support the regeneration and the protection of the native structure.\\
In this paper we focus on the hygroscopic behavior of ectoine and hydroxyectoine in comparison to urea. Special emphasis is spent on the structural ordering of the local water environment. 
These results are important for a further understanding of the mechanisms of compatible solutes and their corresponding functionality.\\ 
We have performed Molecular Dynamics simulations where the specific influence of extremolytes on the molecular structure of water is investigated via radial distribution functions. 
The influence of a varying salt concentration was also investigated.
Our conclusions about the structural alteration of the local solvent environment are supported by experimental results of a DPPC monolayer phase transition.
It has been recently discussed \cite{Galla11} that compatible solutes interact with the membrane via a solvent-mediated mechanism. 
Hence, we will show that the relative influence in the ordering of water by a compatible solute 
is directly reflected in the broadening mechanism of a liquid expanded/liquid condensed phase transition.  
It has to be noticed that we do not aim to numerically investigate this mechanism in detail. The experimental results serve as a validation for the perturbation exerted by different compatible solutes 
on the local water environment as it has been found in our simulations.\\ 
In addition we discuss the hygroscopic properties of the different molecules in atomistic detail. Our results allow to interpret the numerical findings of a recent publication \cite{Yu07} and validate 
the experimental consensus that extremolytes are more appropriate for water binding than other compatible solutes \cite{Lentzen06,Driller08,Dirschka08,Galla10}. 
We will show that the length scale of solvent perturbation is identical to the distance between a protein and a compatible solute in agreement to Ref. \cite{Yu07}.\\ 
The paper is organized as follows. In the next sections we present the theoretical background and the numerical and experimental details. The results and the discussion are presented in the fourth section and  
we finally conclude with a brief summary in the fifth section.
\section{Numerical Details}
Our simulations have been performed in explicit SPC/E water model solvent \cite{Straatsma87} with the software package GROMACS \cite{Berendsen95,Hess08,Spoel05} 
and variations of the force field GROMOS96 \cite{Oostenbrink04}. The initial chemical structure was created via the SMILES language
\cite{Weininger88,html}. The topology file with the essential molecular dynamics
parameters has been derived by the usage of the PRODRG server \cite{PRODRG}.
The structure and the charges of ectoine, hydroxyectoine and urea have been refined and calculated by the software package 
TURBOMOLE 6.1 \cite{TM89}. By applying M\o{}ller-Plesset perturbation theory (RI-MP2) with the TZVPP basis 
set \cite{Ahlrichs98, Haettig05}, we performed a geometry optimization in combination with the COSMO solvent model \cite{Eckert00,Klamt93}. We found that the zwitterionic forms of ectoine and hydroxyectoine are 
10.81 kcal/mol and 12.35 kcal/mol, respectively, more stable than the neutral counterparts. Hence we neglect all Molecular Dynamics simulations regarding neutral ectoine and neutral hydroxyectoine.\\
After geometry optimization and calculation of the atomistic properties, we transferred the corresponding values for the equilibrium parameters of the bond length, the partial charges and the bond angles 
to the PRODRG topology file. The modified PRODRG file, where all other atomistic parameters as given by the 
GROMOS96 force field \cite{Oostenbrink04} were left unchanged, was finally used for the classical Molecular Dynamics simulations.\\
The atomistic simulations have been carried out in a 
cubic simulation box with periodic boundary conditions. The box for ectoine was
$(4.5342 \times 4.5342\times 4.5342)$ nm$^3$ filled with 3077 SPC/E water molecules.  
Hydroxyectoine has been simulated within a box size of   
$(4.6153 \times 4.6153\times 4.6153)$ nm$^3$ and 3261 water molecules. 
The results for urea have been derived in a box of $(4.4015 \times 4.4015\times 4.4015)$ nm$^3$ with 2803 water molecules. The solvent density was nearly identical for all simulations.\\ 
Electrostatic interactions have been calculated by the Particle Mesh Ewald sum \cite{Pedersen95} in an electroneutral system. 
Varying salt concentrations have been achieved by replacing water molecules by the analog number of chloride and sodium ions. \\
The time step was $\delta t=2$ fs and the temperature was kept constant at 300 K by a Nose-Hoover thermostat \cite{Frenkel96}. All bonds have been constrained by the LINCS algorithm \cite{Fraaije97}. 
After minimizing the energy, we performed 100 ps of equilibration followed by a 10 ns simulation sampling run.\\ 
The structure of the solution has been investigated by the radial distribution function \cite{Allen86}
\begin{equation}
g_{_{AB}}(r) = \frac{<\rho_{_B}(r)>}{<\rho_{_B}>}
\label{eq:rdf}
\end{equation}
which gives the probability of finding a molecule $B$ around molecule $A$ compared to the ideal gas description,
where $<\rho_B(r)>$ denotes the particle density of type $B$ at a distance $r$ around particle $A$ and $<\rho_B>$ describes 
the average particle density of $B$ within a distance $d$ around $A$.\\ 
The cumulative particle number function is given by 
\begin{equation}
f_{_{AB}}(r) = 4\pi <\rho_{_B}>\int_{0}^d r^2\; dr \; g_{_{AB}}(r).
\label{eq:frdf}
\end{equation}
and yields an estimate for the number of molecules within the distance $d$.
A further important quantity is the number of present hydrogen bonds $n_{_{HB}}$.  
Division of this value by the total solvent accessible surface area $\sigma_t$ gives
\begin{equation}
\label{eq:HBD}
\rho_{_{HB}}= \left<\frac{n_{_{HB}}(t)}{\sigma_t(t)}\right>
\end{equation}
the hydrogen bond density \cite{Yu07} that counts the average number of hydrogen bonds per unit surface area.
The solvent accessible surface area is calculated by the sum of spheres centered at the
atoms of the studied molecule, such that a spherical solvent molecule can be placed in closest distance and in agreement to van-der-Waals interactions by following the constraint that other atoms are not 
penetrated \cite{Scharf95}.
A hydrogen bond is present, if the distance between the interacting atoms is closer than 0.35 nm and the interaction angle is not larger than 30 degrees.\\ 
To compare the molecules we have calculated the solvent accessible hydrophilic surface area $\sigma_h$ \cite{Scharf95} as given for the hydrophilic atoms of the molecule, which is 
divided by the total solvent accessible surface area resulting in the ratio $r_{\sigma}$.\\
The influence of the charged functional groups can be estimated by conducting simulations where the partial charges of the compatible solutes have been set to zero such that the complete molecule was uncharged. 
The number of hydrogen bonds calculated in these simulations were
subtracted from the average number of hydrogen bonds $n_{_{HB}}$ present at the charged molecules resulting 
in the net average number of hydrogen bonds $n_{_{HB}}^{c}$.
The difference of the charged and uncharged molecules cumulative radial distribution function of water molecules is given by $\Delta f(r)=f_{_{AB}}(r)-f_{_{AB}}^{uc}(r)$.
\section{Experimental details}
The lipid DPPC (1,2-dipalmitoyl-sn-glycero-3-phosphocholine) was purchased from Avanti Polar Lipids Inc. (Alabaster, AL). 
2-(4,4-Difluoro-5-methyl-4-bora-3a,4a-diaza-s-indacene-3-dodecanoyl)-1-hexadecanoyl-sn-glycero-3-phosphocholine ($\beta$-BODIPY$^{\small{\textregistered}}$ 
500/510 C$_{12}$-HPC, BODIPY-PC) was obtained from Molecular Probes (Eugene, OR). 
The DPPC was dissolved in chloroform/methanol solution (1:1, v/v). Chloroform and methanol were high pressure liquid chromatography grade and purchased from Sigma-Aldrich (Steinheim, Germany) 
and Merck (Darmstadt, Germany), respectively. 
Urea was purchased from Roth Chemie GmbH (Karlsruhe, Germany). 
Ectoine ((S)-2-methyl-1,4,5,6-tetrahydropyrimidine-4-carboxylic acid) and hydroxyectoine ((4S,5S)-2-methyl-5-hydroxy-1,4,5,6-tetrahydropyrimidine-4-carboxylic acid) were obtained from Bitop AG (Witten, Germany). 
Water was purified and deionized by a multicartridge system (MilliPore, Billerica, MA) and had a resistivity $>18$ M$\Omega$m.\\
Domain structures of DPPC doped with 0.5 mol\% BODIPY-PC were visualized by means of video-enhanced epi-fluorescence microscope (Olympus STM5-MJS, Olympus, Hamburg, Germany) equipped with a 
xyz-stage and connected to a CCD camera (Hamamatsu, Herrsching, Germany). 
The analytical Wilhelmy film balance (Riegler and Kirstein, Mainz, Germany) with an operational area of 144 cm$^2$ was placed on the xyz-stage of the microscope.
All measurements were performed on the subphase containing pure water, ectoine, hydroxyectoine or urea, at 20$^\circ$ C.
The DPPC solution from an organic solvent was spread on the subphase and left for $10-15$ min for the solvent to evaporate followed by compressing the monolayer at a rate of 2.9 cm$^2$/min.\\
The images were captured by stopping the barrier and equilibrating the monolayer for few minutes at desired surface pressures.
\section{Results and discussion}
To understand the influence of compatible solutes on an aqueous solution, we analyzed the configuration of water molecules around ectoine, hydroxyectoine and urea and studied several static properties. 
\begin{table}[h!]
\caption{Hydrophilic surface $\sigma_{h}$ and its ratio $r_{\sigma}$ to the total surface area, number of hydrogen bonds $n_{_{HB}}$, net number of hydrogen bonds $n_{_{HB}}^{c}$ 
and average hydrogen bond density $\rho_{HB}$.}
\label{tab:2}       % Give a unique label
\begin{tabular}{llll}
\hline\noalign{\smallskip}
Molecule & Urea & Ectoine & Hydroxyectoine \\
\noalign{\smallskip}\hline\noalign{\smallskip}
$\sigma_h$ [nm$^2$] & $1.41 \pm 0.01 $ & $1.63 \pm 0.01$ & $1.73 \pm 0.01$\\
\noalign{\smallskip}\hline\noalign{\smallskip}
$r_{\sigma}$ & 1 & $0.64\pm 0.01$ & $0.67\pm 0.01$\\
\noalign{\smallskip}\hline\noalign{\smallskip}
$n_{_{HB}}$ & 4.43 & 7.01 & 8.77\\
\noalign{\smallskip}\hline\noalign{\smallskip}
$n_{_{HB}}^{c}$ & 3.35 & 6.31 & 7.75\\
\noalign{\smallskip}\hline\noalign{\smallskip}
$\rho_{_{HB}}$ [nm$^{-2}$] & $3.14\pm 0.03$ & $2.75 \pm 0.01$ & $3.39 \pm 0.02$\\
%$\rho_{_{HB}}^h$ [nm$^{-2}$] & $3.14\pm 0.03$ & $4.30 \pm 0.01$ & $5.07 \pm 0.02$\\
\noalign{\smallskip}\hline
\end{tabular}
\end{table}
The results are presented in Tab.~\ref{tab:2}. 
It is obvious that the ectoines are different compared to urea in their ratio of the hydrophilic to the total surface area $r_{\sigma}$. The hydrophilic surface area $\sigma_h$ 
shows a slight increase in the order of urea, ectoine and hydroxyectoine. The net number of hydrogen bonds $n_{_{HB}}^c$ is larger for ectoine and hydroxyectoine. The values for $n_{_{HB}}^c$ are slightly smaller 
than $n_{_{HB}}$ for all molecules. 
The hydrogen bond densities for the total surface areas $\rho_{_{HB}}$ are in the range of $2.75 - 3.39$ nm$^{-2}$.
Hence compared to the values of $n_{_{HB}}$, it is obvious that the number of hydrogen bonds constantly grows with the total surface area. Thus the larger number of hydrogen bonds for ectoine and hydroxyectoine can 
be partly related to a simple size effect.\\ 
\begin{figure}[h!]
\includegraphics[width=0.5\textwidth]{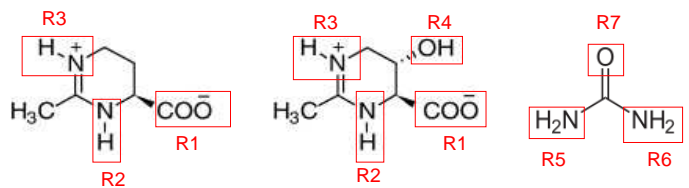}
\caption{Polar groups in ectoine (left), hydroxyectoine (middle) and urea (right).}
\label{fig:4}
\end{figure}
\begin{table}
\caption{Average hydrogen bonds and average lifetime per functional group (in brackets) per molecule.}
\label{tab:3}       % Give a unique label
\begin{tabular}{llll}
\hline\noalign{\smallskip}
Group & Ectoine & Hydroxyectoine & Urea \\
\noalign{\smallskip}\hline\noalign{\smallskip}
R1 & 5.59 (11.99 ps) & 5.26 (13.08 ps)& -- \\
\noalign{\smallskip}\hline\noalign{\smallskip}
R2 & 0.68 (5.81 ps) & 0.84 (7.78 ps) & -- \\
\noalign{\smallskip}\hline\noalign{\smallskip}
R3 & 0.75 (6.89 ps)  & 0.69 (6.32 ps) & -- \\
\noalign{\smallskip}\hline\noalign{\smallskip}
R4 & -- & 1.98 (9.89 ps) & -- \\
\noalign{\smallskip}\hline\noalign{\smallskip}
R5 & -- & -- & 0.90 (5.24 ps) \\
\noalign{\smallskip}\hline\noalign{\smallskip}
R6 & -- & -- & 0.83 (5.35 ps) \\
\noalign{\smallskip}\hline\noalign{\smallskip}
R7 & -- & -- & 2.71 (13.40 ps) \\
\noalign{\smallskip}\hline
\end{tabular}
\end{table}
To investigate the characteristics of the water binding properties in more detail, we further studied the effects of the functional groups on the corresponding number of present hydrogen bonds. 
Therefore we identified the polar groups in each molecule and numbered them by R1-R7 (Fig.~\ref{fig:4}). 
The data together with the
corresponding average lifetimes are displayed in Tab.~\ref{tab:3} with the notation given in Fig.~\ref{fig:4}.\\ 
Most of the hydrogen bonds are present at the carboxylic group (R1) of 
the extremolytes. An increased value of hydrogen bonds for urea is induced by the oxygen group (R7). The lifetimes for a hydrogen bond with urea (R7) and for the extremolytes (R1) are comparable.
The average lifetime of hydrogen bonds between water molecules has
been found to be around $2.72$ ps which indicates that the values displayed in Tab.~\ref{tab:3} are induced by well coordinated and localized water molecules.
Regarding the results for the hydrogen bond density $\rho_{_{HB}}$ and the number of hydrogen bonds $n_{_{HB}}$ in Tab.~\ref{tab:2}, 
it becomes clear that a non-uniform accumulation of water molecules around the ectoines is mainly induced by the carboxylic group (R1).\\
By the analysis of these properties, we have shown that a stronger hygroscopic effect is mainly induced by the presence of zwitterionic groups compared to urea. Additionally the size of the molecule 
is of importance due to the total solvent accessible surface area in contrast to the ratio of the hydrophilic surface area. Thus the pure static properties imply an advantage of the ectoines 
over urea which is less able to bind water molecules. Hence water capturing at dry situations is better achieved by the extremolytes in agreement to their biological function. 
\begin{figure}[h!]
 \includegraphics[width=0.5\textwidth]{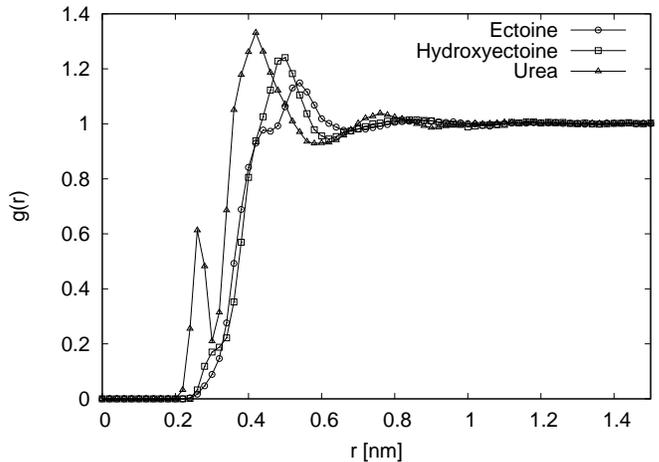}
\caption{Radial distribution function $g(r)$ for water molecules around the compatible solutes
ectoine, hydroxyectoine and urea. }
\label{fig:fig6}
\end{figure}
The results for the radial distribution function of water molecules around the center of mass for the compatible solutes are presented in
Fig.~\ref{fig:fig6}. Hydroxyectoine and ectoine have nearly identical distribution functions with pronounced peaks at 0.5 nm (hydroxyectoine) and 0.54 nm (ectoine).
Urea shows two peaks at 0.24 nm and 0.4 nm. This can be explained by the size of urea which allows the water molecules to form a hydration sphere on a smaller length scale.
The influence on the molecular water structure decays around 1 nm to
reach the bulk value for all molecules. Hence the main influence of the compatible solutes on the water molecules is located within this distance.\\
A more detailed investigation of the water binding to the functional groups has been also performed (data not shown). 
The carboxylic group R1 is able to form a pronounced structuring of the water molecules around distances 0.22 and 0.34 nm. 
The main contribution at 0.54 nm for ectoine and 0.5 nm 
for hydroxyectoine as it was shown in Fig.~\ref{fig:fig6} is related to the effect of all contributing functional groups. 
Hence at these distances all polar groups are able to bind water molecules which yield a concerted contribution to 
the radial distribution function resulting in the peaks presented in Fig.~\ref{fig:fig6}. Additionally it can be shown that the large peak for urea at 0.2 nm is mainly caused by the presence of oxygen (R7). 
Thus it can be concluded that all studied molecules are able to 
form a pronounced accumulation sphere of water molecules at short distances. With these results we are able to indicate the length scale of the most significant interactions. 
As it has been discussed in Refs.\cite{Yu07,Galla11}, extremolytes interact with macromolecules via solvent mediated effects. Hence relevant interactions have to be taken place within this distance.\\ 
The results for the radial distribution function are in good agreement to the results for the local pressure of the water molecules around the compatible solutes 
shown in Fig.~\ref{fig:fig15} which was calculated by the method proposed in Ref.~\cite{Ollila09}. In general the values for the pressure are mainly dominated by interparticle forces.
It can be shown \cite{Ollila09} that the pressure tensor $p_{\alpha\beta}(r)$ depends on $p_{\alpha\beta}(r)\sim F_{ij,\alpha}r_{ij,\beta}$ where $F_{ij}$ denotes the conservative interparticle force and 
$r_{ij}$ the distance between the particle $i,j$ along the directions $\alpha,\beta$. Hence it is obvious, that negative parts of the pressure correspond to attractive forces of the water molecules 
towards the center of the compatible solutes 
and positive values can be related to repulsive interactions.
Large negative values can be observed for urea on distances up to 0.45 nm indicating strong attractive interactions. Hydroxyectoine and ectoine have negative 
values on distances up to 0.55 and 0.6 nm, respectively. It is obvious that the local pressure on scales 0.45-0.55 nm is drastically decreased in presence of ectoine. Additionally it can be seen that 
hydroxyectoine
has a direct influence on the local solvent pressure by a large reduction on scales 0.425 to 0.525 nm. Hence, it can be concluded that water molecules are strongly correlated with the compatible solutes 
on short length scales which is significantly indicated by the discussed 
negative values. It is obvious that these results are in good agreement to Fig.~\ref{fig:fig6} although a detailed analytical derivation is complicated \cite{Allen86}. 
Additionally we have calculated the pressure differences between charged and uncharged compatible solutes.
It comes out that the differences additionally show an attractive net interaction between the compatible solutes and the water molecules as presented in the Supplementary material. It can be assumed that
in presence of a protein or a monolayer, the strong interaction towards the center of the compatible solutes causes a significant contribution on the hydration shell of the macromolecule which can be 
additionally responsible for the preservation of the native state. This is in particular important for distances between the compatible solute and the protein which are smaller than 0.7 nm as they 
have been reported in Ref.\cite{Yu07}.\\
For a further discussion, we analyzed the relevant length scales for all molecules which can be related to different types of interactions.
Very short distances which are smaller than 0.4 nm may correspond to direct interactions like hydrogen bonds. 
In addition, on a longer length scale $r>0.8$ nm only small variations can be observed in Fig.~\ref{fig:fig15}. Hence it can be again concluded that the relevant length scale $r=0.4-0.6$ nm is the
most important one for the influence of compatible solutes on other molecules as it has been mentioned in the introduction and in Ref.~\cite{Yu07}.
The drastic influence on the local water
pressure in the important interval $r=0.4-0.6$ nm is obvious.
Hydroxyectoine has an average pressure of -4200 bar, ectoine a value of -3200 bar and urea a positive value of 550.\\
\begin{figure}[h!]
 \includegraphics[width=0.5\textwidth]{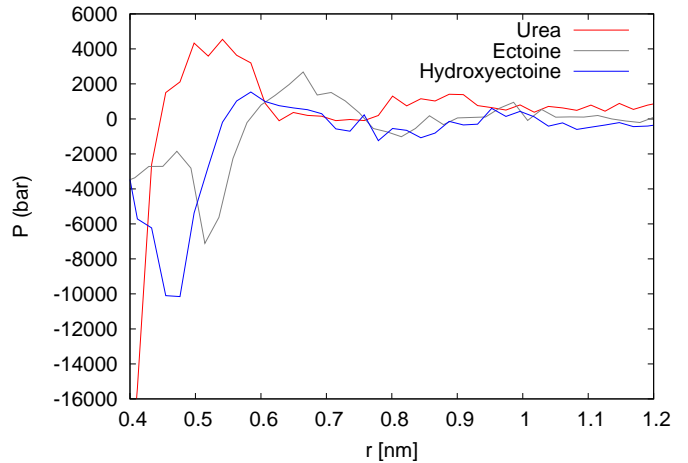}
\caption{Pressure in the bulk water phase around the compatible solutes.}
\label{fig:fig15}
\end{figure}
The effect of water accumulation can be also considered by calculating the radial cumulative number distribution function after Eqn.~\ref{eq:frdf}. 
For a comparison, we conducted simulations with uncharged compatible solutes by setting all partial charges to zero. The results for the distribution function derived by these simulations are used to define 
a reference system which aims to investigate the influence of the atomic charges. 
\begin{figure}[h!]
 \includegraphics[width=0.5\textwidth]{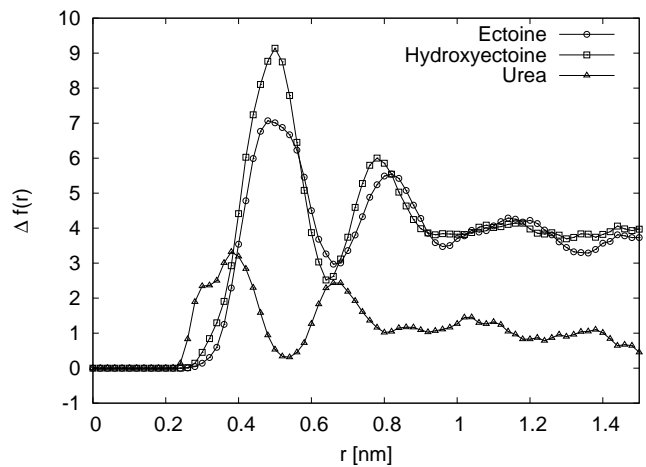}
\caption{Difference of the radial cumulative number distribution function $\Delta f(r)$ for water molecules around the compatible solutes.}
\label{fig:fig9}
\end{figure}
Our findings for the differences of the radial cumulative number distribution function $\Delta f(r)$ are shown in Fig.~\ref{fig:fig9}. It is evident that hydroxyectoine is able 
to bind up to nine water molecules, respectively seven water molecules for ectoine on a distance smaller than 0.5 nm. Urea binds approximately three molecules on a distance $r<0.4$ nm. 
These results are closely related to the average number of hydrogen bonds displayed in Tab.~\ref{tab:3} and are in agreement to the hygroscopic characteristics of the molecules. 
The larger values for hydroxyectoine in contrast to ectoine are caused by 
the hydroxyl group (R4) which is able to bind nearly two water molecules (Tab.~\ref{tab:3}).
However, a significant effect can be observed on distances larger than 1 nm. The net value for accumulated water is given by approximately four molecules for the ectoines and one for urea. Hydrogen bonding
is not presented at these distances which indicates that these water molecules are effectively accumulated due to indirect ordering and electrostatic effects.
Thus the indirect long range influence of extremolytes on the local water structure is 
obvious. We conclude from these data that extremolytes are able to significantly change the local water environment. Due to the analysis of the relevant length scales, one can assume that these perturbations
are mainly responsible for the preservation mechanism of macromolecules. 
\subsection{Influence of a varying salt concentration} 
Another interesting quantity is the direct influence of salt on the hygroscopic properties of the solutes. 
As it was mentioned in the introduction, extremolytes allow the cell to resist high salinity. 
Therefore we expect the influence of salt to be negligible for the properties of the extremolytes.\\
To investigate this point in more detail, we simulated salt solutions with concentration values ranging from 0.017 mol/L to 0.47 mol/L which is comparable to
experimental and physiological conditions \cite{Lentzen06}. 
As it can be concluded from the data shown above,
main indicators for deliquescent properties are large numbers of apparent hydrogen bonds. 
\begin{figure}[h!]
 \includegraphics[width=0.5\textwidth]{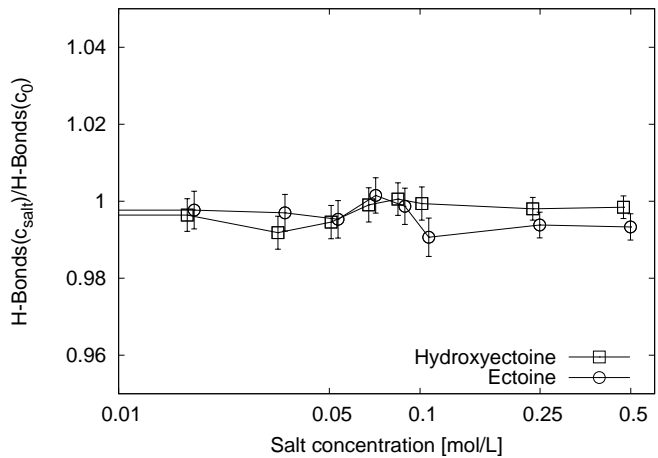}
\caption{Number of hydrogen bonds in presence of salt divided by the average values without salt given in Tab.~\ref{tab:2}}.
\label{fig:fig11}
\end{figure}
Therefore we have compared the number of hydrogen bonds for an increasing salt concentration to the ratio of the values for 
the zero salt simulations given in Tab.~\ref{tab:2}. The results are shown in Fig.~\ref{fig:fig11}.\\
The ratio is always above 0.98 indicating a minor influence of salt on the number of hydrogen bonds.
Therefore it can be concluded that high salinity does not affect the hygroscopic properties of extremolytes. This implies that the extremolytes are able to bind water molecules
even under harsh conditions which allows to protect the cell against drying out. 
\subsection{Effects of compatible solutes on the structural organization of DPPC monolayers}
\begin{figure}[h!]
 \includegraphics[width=0.5\textwidth]{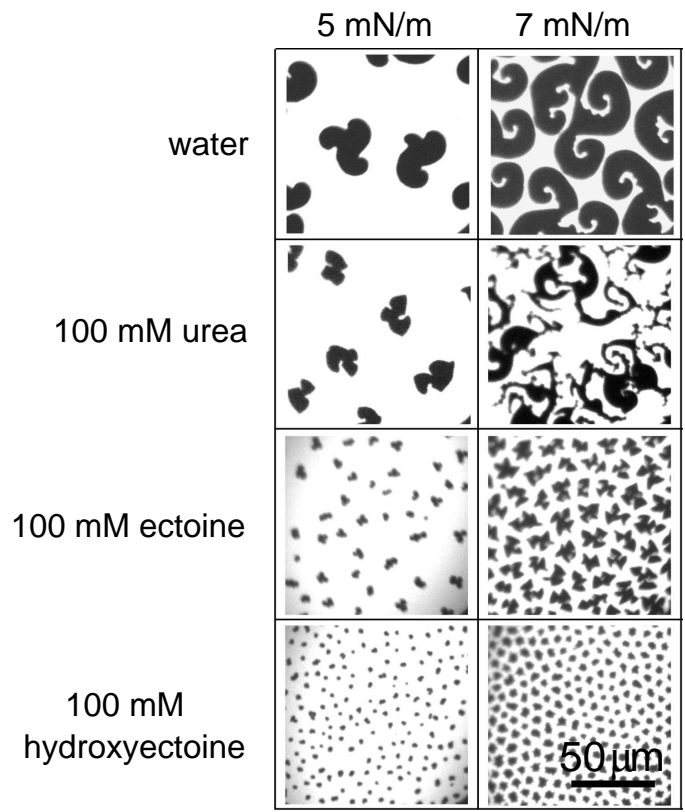}
\caption{Video-enhanced fluorescence microscopic images of DPPC monolayers on subphases containing ectoine, hydroxyectoine and urea, All measurements were performed at 20 $^{\circ}$C and surface pressures 
5 (left) and 7 mN/m (right). The results for hydroxyectoine and ectoine have been taken from \cite{Galla10} for the ease of comparison.}
\label{fig:fig14}
\end{figure}

As we have mentioned in the introduction an important impact of extremolytes on biomolecular assemblies is
to increase the fluidity of membranes under specific circumstances \cite{Galla10,Galla11}. 
Regions with increased flexibility are included in the liquid expanded (LE) phase whereas more rigid domains are given in the liquid condensed (LC) phase.\\
In a lipid monolayer the formation of several domains can be experimentally observed under certain temperature conditions in coexistence of the LE phase \cite{Vanderlick1997, Grimellec03}. 
A driving force to understand the formation of these 
regions has been indicated to be the line tension \cite{McConnell92,May00}.\\ 
Fluid phase coexistence in lipid membranes is characterized by the formation of both LC phase as well as the LE phase. 
The LC phase exhibits a higher degree of ordering and packing preventing water to penetrate the hydrophobic core. 
Hence the mechanical elasticity is increased in the LE phase. It has been shown \cite{Galla10} that the presence of hydroxyectoine and ectoine modifies the formation of the LC domains and therefore 
ensures the fluidization of
the monolayer. This effect has been described by a variation of the preferential exclusion model \cite{Lee81,Arakawa83,Arakawa85,Yu04,Yu07,Galla10} as well as a direct influence of the extremolytes on the 
lipid headgroup region \cite{Galla11}.\\ 
We have carried out experiments for a DPPC monolayer in presence of the different compatible solutes hydroxyectoine, ectoine and urea to investigate the influence of each species on the LE/LC phase transition. 
Due to the relative ordering of the molecules by their hygroscopic properties, we expect the phase transition to reflect this behavior. \\ 
The fluorescence pictures of the monolayers doped with BODIPY-PC, a fluorescent dye that is preferentially soluble in the fluid LE phase are shown in Fig.~\ref{fig:fig14}. 
Dark domains represent the rigid LC phase, light areas the more fluid LE phase. Lipid films on pure water exhibit the well known kidney shaped LC domains in the phase transition 
region \cite{Galla10}. The extent of the LC phase increases for all molecules with increasing surface pressure.\\  
In pure water the characteristic multi-lobed structure starts to form at 5 mN/m and is fully pronounced at 7 mN/m as a left handed clearly developed nanostructure. 
The presence of urea is only important at 7 mN/m where the domain size
is considerably decreased in comparison to the results of pure water. No visible effect can be observed at 5 mN/m.\\ 
Ectoine at 5 mN/m leads to a considerable shrinkage in the domain size. 
At both 5 mN/m and 7 mN/m the well developed domain structure is therefore lost continuously. This effect is even more enhanced for a solution containing hydroxyectoine.\\
Thus it is obvious that hydroxyectoine and ectoine have a more distinct effect than urea on DPPC monolayers which is in accordance to the numerical results shown above.
It can be assumed that 
the effects appear on a length scale larger
than 0.4 nm \cite{Yu07} in front of the macromolecular surface which sheds a new light on the second order packing structures indicated by Fig.~\ref{fig:fig9}.\\ 
As it is obvious and discussed above, the extremolytes have a net effect on long and short distances 
on the solvent configuration. It can be concluded that the specific alteration of the solvent environment as indicated by the results shown above is directly reflected in the broadening of the monolayer phase transition. 
Hence the experimental results validate the relative ordering of the compatible solutes related to their hygroscopic characteristics.
These results are also in good agreement to the observed surface pressure isotherms that are presented in the supplementary material.
\section{Summary and conclusion}
We have investigated the microscopic deliquescent properties of the compatible solutes and their influence on the local water structure.
Our results indicate that ectoine and hydroxyectoine 
are able to bind more water molecules than urea. Thus the usage of extremolytes as water binding chemical detergents in dermatologic products is validated.
By comparison to the uncharged, purely steric interacting species 
we found that the extremolytes are able to accumulate nine and seven water molecules for hydroxyectoine and ectoine, respectively, in a distance located within 0.6 nm. The reason
for this behavior is the formation of a large number of hydrogen bonds at specific functional groups of the molecules.\\ 
Furthermore we have determined a net value of four accumulated water molecules for the extremolytes on length scales larger than 0.6 nm. This value can be related to second order packing effects 
due to the fact that direct interactions like hydrogen bonds are absent on these distances.\\
Although the extremolytes are charged and act via electrostatic interactions with the solvent, we have shown that the presence of salt does only slightly alter the 
number of hydrogen bonds. Hence, the specific water binding behavior of the compatible solutes is not perturbed at high salt concentrations.
Compared to urea, ectoine and hydroxyectoine have more distinct hygroscopic properties. 
One can therefore identify extremolytes as water-binding molecules which order the local structure of the solvent on a length scale of 0.6 nm significantly. On longer length scales 
second order effects can be observed which result in net water accumulation due to packing effects.\\
Finally we experimentally studied the impact of the compatible solutes on the formation of domains in the LE-LC phase transition of lipid monolayers. Our results indicate that ectoine and hydroxyectoine
broaden the phase transition by suppressing the appearance of LC domains. Thus the mechanical flexibility of the monolayer is more enhanced in the presence of extremolytes compared to urea. 
These results can be brought into accordance to 
the numerical data concerning the relative hygroscopicity and the long range ordering effects of the local water structure.\\
Summarizing all results, our study has shown that the hygroscopic properties of several compatible solutes significantly differ in their specific behavior. 
Our findings allow to achieve a more detailed view on the structural alteration of the local water environment. As it has been discussed in the introduction, the stabilizing effect of co-solvent on proteins and 
lipid monolayer has been studied in detail before. However, a theoretical investigation concerning the important length scales has been missing.\\
We have shown that the relevant distances between interacting species which have been found in the literature at 0.4-0.6 nm \cite{Yu07} are validated by a significant modification
of the water ordering. Hence, our results allow to conclude that this distance is naturally implied by the effects of solvent ordering. Although we have neglected the presence of target molecules, our results indicate 
the large hygroscopicity of the ectoines as a main reason for the preservation mechanism. Due to the zwitterionic properties and the hydrophobic/hydrophilic surface area contributions, 
the ectoines are repelled from a proteins surface such that the deliquescent characteristics lead to a strong binding of water molecules. It can be assumed that some parts of these 
processes are mainly involved in the stabilization mechanism.\\
Due to the actual large number of hydrogen bonds even in high salt concentrations, we have validated that a significant increase of ions does not change the molecular properties.
Electrostatic screening of the zwitterionic parts can be neglected such that the functionality of the co-solvents is conserved even under harsh conditions. Thus, we have validated 
that the ectoines are a specific class of molecules which significantly differ from other osmolytes due to their hygroscopic characteristics and the variation of the water environment.
Strong binding of solvent molecules indicates them as perfect molecules against drying out.
The evolutionary strategy to overcome dry situations for microorganisms by the synthesis of extremolytes is supported by our data. All our results indicate 
that ectoine and hydroxyectoine are better suited as hygroscopic detergents in dermatological products than long known hygroscopic molecules like urea. 
\section{Acknowledgments}
Financial support by the Deutsche Forschungsgemeinschaft (DFG) through the transregional collaborative research center TRR 61 and the SFB 858 is gratefully acknowledged. 
R. K. Harishchandra acknowledges funding from the International Graduate School in Chemistry at the University of M{\"u}nster. Additional results can be found in the supplementary material. 

\end{document}